\documentclass[letterpaper]{article}

\usepackage[letterpaper,top=2cm,bottom=2cm,left=3cm,right=3cm,marginparwidth=1.75cm]{geometry}

\usepackage{amsmath}
\usepackage{graphicx}
\usepackage[colorlinks=true, allcolors=blue]{hyperref}
\usepackage{amssymb}
\usepackage{booktabs}
\usepackage{multirow}
\usepackage{xcolor}
\usepackage{parskip}
\usepackage{comment}
\usepackage[numbers, sort]{natbib}
\usepackage{authblk}

\usepackage{xeCJK}
\newcommand{\ja}[1]{#1}

\makeatletter
\newcommand{\@toptitlebar}{%
  \hrule height 4\p@
  \vskip 0.25in
  \vskip -\parskip%
}
\newcommand{\@bottomtitlebar}{%
  \vskip 0.29in
  \vskip -\parskip
  \hrule height 1\p@
  \vskip 0.09in%
}

\renewcommand{\@maketitle}{%
  \vbox{%
    \hsize\textwidth
    \linewidth\hsize
    \vskip 0.1in
    \@toptitlebar
    \centering
    {\fontsize{17pt}{20pt}\selectfont\bfseries \@title\par}
    \@bottomtitlebar
    
    \vskip 0.2in
    \let\footnote\thanks
    {\large \@author\par}
    \vskip 0.3in minus 0.1in
  }
}
\makeatother

\definecolor{cmzhao}{rgb}{0.1, 0.8, 0.1}
\definecolor{zhu}{rgb}{0.6, 0.2, 0.2}

\title{Sarashina2.2-TTS: Tackling Kanji Polyphony in Japanese Speech Generation via Data Scaling and Targeted Data Synthesis}

\author{Lianbo~Liu}
\author{Shiao~Zhu}
\author{Kai~Washizaki}
\author{Reo~Yoneyama\footnote{This work was conducted during his internship at SB Intuitions.}}
\author{Haesung~Jeon}
\author{Mengjie~Zhao}
\author{Yusuke~Fujita}
\author{Hao~Shi\footnote{Work done while at SB Intuitions.}}
\author{Nao~Yoshida}
\author{Yuan~Gao}
\author{Roman~Koshkin}
\author{Yukiya~Hono}
\author{Yui~Sudo}

\affil{SB Intuitions \\ \text{lianbo.liu@sbintuitions.co.jp}}

\date{}

\begin{document}

\maketitle


\begin{abstract}
\noindent
While large language model (LLM)-based text-to-speech (TTS) systems have achieved high-quality speech synthesis, most existing systems focus on English and Chinese. 
Japanese, however, remains under-explored, and its unique linguistic challenges, such as widespread context-dependent kanji polyphony, have yet to be adequately tackled.
Here we introduce Sarashina2.2-TTS\footnote{Model weights and code are available at \url{https://github.com/sbintuitions/sarashina2.2-tts}}, a Japanese-centric LLM-TTS system that tackles these challenges through a dual approach: data strategy and evaluation methodology. 
First, we scale training to approximately 361k hours of speech, incorporating a balanced mix of Japanese and English data. Furthermore, we design a targeted data augmentation pipeline covering all 2,136 Joyo (regular-use) kanji designated by Japan's Agency for Cultural Affairs to efficiently address kanji polyphony disambiguation.
Second, we introduce the Joyo Kanji Yomi Benchmark\footnote{Code and data are available at \url{https://github.com/sbintuitions/JoyoKanji-Yomi-Benchmark}}, covering all 2,136 Joyo kanji and their 4,378 readings. Alongside this benchmark, we propose Kana-CER, a metric that compares synthesized speech against reference readings in the kana space, eliminating orthographic variations to directly measure pronunciation correctness. 
Experiments demonstrate that our targeted data augmentation significantly improves reading accuracy. Overall, Sarashina2.2-TTS achieves state-of-the-art kanji-level reading accuracy and matches top baselines on general sentence-level pronunciation, while delivering the highest speaker similarity in zero-shot Japanese speech synthesis. Furthermore, cross-lingual evaluation reveals that Sarashina2.2-TTS is the only system that maintains stable Japanese pronunciation regardless of the prompt language, 
confirming that our balanced training approach improves cross-lingual robustness.
\end{abstract}

\section{Introduction}

Recent text-to-speech (TTS) systems built on large language models (LLMs) achieve highly natural and expressive speech synthesis by leveraging large-scale training datasets \citep{cosyvoice3, seedtts, qwen3tts, fireredtts2}. However, most existing LLM-TTS efforts focus on high-resource languages such as English and Chinese. Although some multilingual systems include Japanese as a supported language \citep{qwen3tts, xtts}, they have not been optimized for the specific linguistic challenges of Japanese, and their performance on Japanese synthesis often remains unsatisfactory.

Japanese poses unique difficulties for TTS, the most critical of which is kanji polyphony. Japanese text interleaves kanji (logographic characters) with kana (phonographic characters). Unlike kana, which uniquely determine pronunciation, the vast majority of kanji have multiple possible readings that are highly dependent on the surrounding context. The 2,136 kanji in the official Joyo Kanji List (List of Regular-Use Kanji)\footnote{\url{https://www.bunka.go.jp/kokugo_nihongo/sisaku/joho/joho/kijun/naikaku/kanji/index.html}} collectively have 4,378 recognized readings, with some kanji having over ten distinct readings. For example, the kanji \ja{``生''} alone has 12 readings including \textit{sei}, \textit{shou}, \textit{ikiru}, and \textit{nama}, each determined by the surrounding context. This makes kanji polyphony disambiguation the central challenge of Japanese TTS.

\paragraph{Challenges.}
We identify two dimensions where current TTS systems fall short: data strategy and evaluation methodology. On the data side, existing multilingual systems that support Japanese suffer from two limitations:
\begin{enumerate}
    \item They typically allocate only a small fraction of their training data to Japanese, providing insufficient exposure to the language's diverse vocabulary and prosodic patterns.
    This data imbalance not only compromises basic Japanese pronunciation accuracy but also severely degrades robustness against cross-lingual prompts.
    \item They lack data engineering strategies specifically designed for kanji polyphony, such as ensuring coverage of diverse kanji readings, particularly infrequent ones that are underrepresented in natural speech corpora.
\end{enumerate}

On the evaluation side, accurately measuring kanji disambiguation itself poses challenges:
\begin{enumerate}
    \item Current benchmarks and metrics lack kanji-level annotations. They can detect that a sentence was mispronounced, but cannot attribute the error to a specific kanji or identify which reading was incorrectly selected, making it impossible to systematically diagnose polyphony errors or guide targeted improvement.
    \item Standard character error rate (CER) and word error rate (WER), computed by comparing text transcribed by an automatic speech recognition (ASR) model against the reference, are confounded by Japanese orthographic variation: the same pronunciation can be written in multiple character forms (e.g., \ja{``行う''}, \ja{``おこなう''}, \ja{``行なう''}), causing spurious errors unrelated to actual pronunciation quality. As a result, Japanese consistently appears as an outlier in multilingual TTS evaluations, with reported CER or WER substantially higher than for other languages \citep{cosyvoice2, xtts, qwen3tts}.
\end{enumerate}

\paragraph{Our approach.}
In this work, we present Sarashina2.2-TTS, a Japanese-centric LLM-TTS system that addresses the above challenges from both sides.

On the data side:
\begin{enumerate}
    \item We train on approximately 361k hours of speech data (194k hours of Japanese across multiple domains and 167k hours of English), the largest Japanese speech dataset used by any open-source TTS system to our knowledge, providing broad vocabulary diversity and prosodic coverage for robust kanji disambiguation.
    \item We propose a targeted synthetic data augmentation pipeline to comprehensively cover infrequent kanji readings underrepresented in natural speech data. To achieve this, we construct a dedicated data synthesis framework that integrates LLM-based sentence generation, dictionary-based prosody annotation, and our newly designed text-side pronunciation control model, Pronunciation Steering (PronSteering).
\end{enumerate}

On the evaluation side:
\begin{enumerate}
    \item We construct the \textbf{Joyo Kanji Yomi Benchmark}, covering all 2,136 regular-use kanji and their 4,378 readings with 13,095 native-speaker-verified test sentences with sentence and kanji level annotations, enabling systematic evaluation and fine-grained error attribution at the kanji level.
    \item We propose \textbf{Kana-CER}, a kana-based character error rate that compares synthesized speech against reference readings in the kana space, eliminating orthographic variation and directly measuring pronunciation correctness.
\end{enumerate}

Experiments demonstrate that Sarashina2.2-TTS outperforms all baselines across all CER-based metrics on the Joyo Kanji Yomi Benchmark and matches the best baseline's pronunciation accuracy on the JSUT\citep{jsut} benchmark, while achieving the highest speaker similarity in zero-shot Japanese speech synthesis. Furthermore, cross-lingual evaluation reveals that Sarashina2.2-TTS is the only system that maintains stable Japanese pronunciation regardless of the prompt language, with virtually no degradation when switching from Japanese to non-Japanese prompts.

\section{Sarashina2.2-TTS} \label{sec:model}

\begin{figure}[h]
    \centering
    \includegraphics[width=0.9\textwidth]{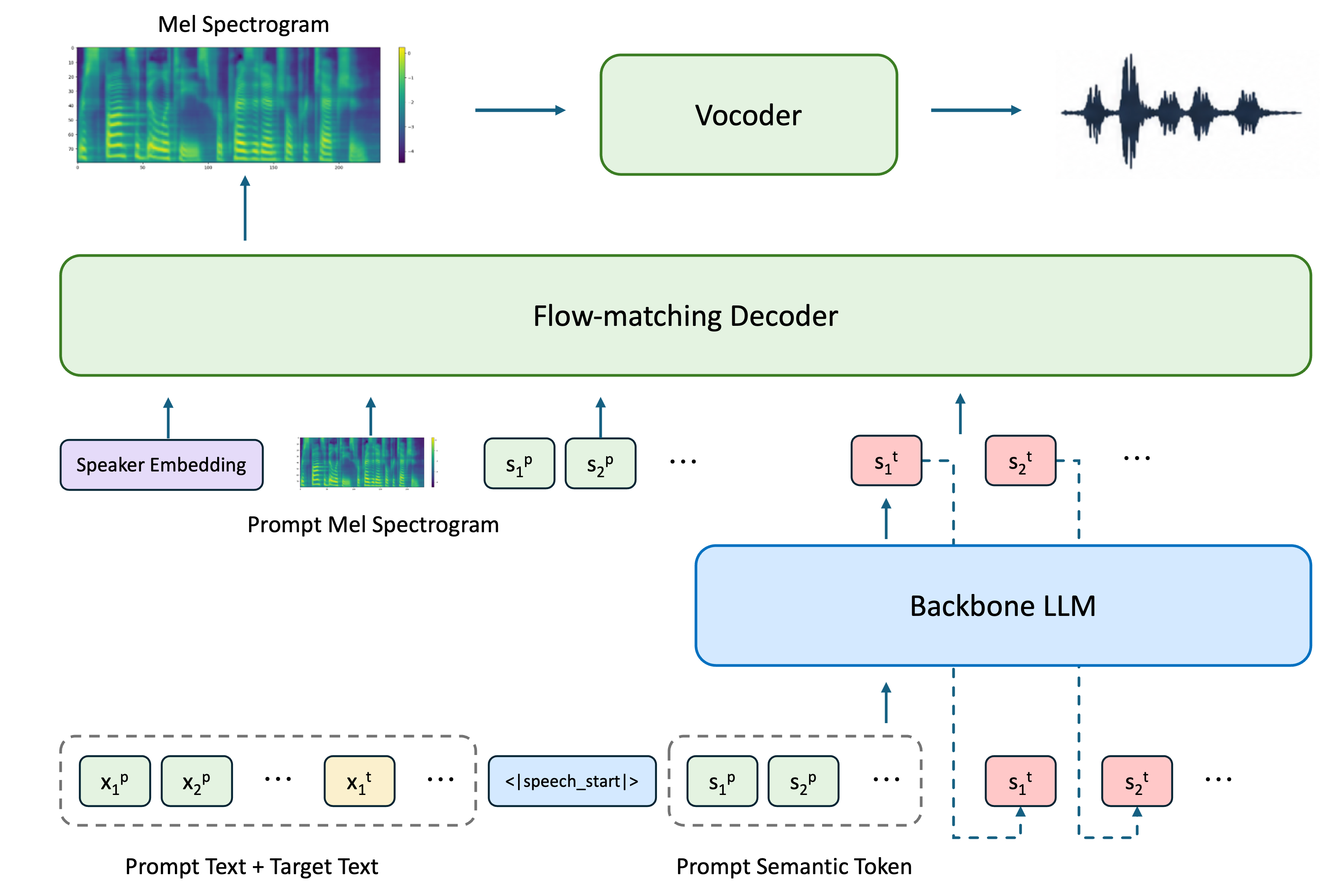}
    \caption{Sarashina2.2-TTS architecture. 
    The semantic stage (backbone LLM) autoregressively maps the prompt and target text, together with prompt semantic tokens from the reference speech, to semantic tokens. The acoustic stage (flow-matching decoder + vocoder) reconstructs the waveform from these semantic tokens, conditioned on the prompt semantic tokens, a speaker embedding and a reference mel-spectrogram.
    }
    \label{fig:architecture}
\end{figure}

Following recent LLM-TTS systems \citep{cosyvoice3, seedtts}, we split speech generation into a semantic stage (the blue block in Figure~\ref{fig:architecture}) and an acoustic stage (the green blocks in Figure~\ref{fig:architecture}). In the semantic stage, a decoder-only backbone LLM autoregressively generates a sequence of discrete semantic tokens, 
conditioned on a concatenated sequence of prompt text, target text, and prompt semantic tokens discretized from reference speech by a speech tokenizer.
In the acoustic stage, a flow-matching decoder takes the semantic tokens together with a speaker embedding and a reference mel-spectrogram as conditions and produces a mel-spectrogram, which is then converted to a waveform by a vocoder. 
These prompt-derived conditions across both stages enable zero-shot voice cloning.
This separation allows the backbone LLM to focus its capacity on context-dependent linguistic decisions, particularly kanji polyphony disambiguation, while acoustic detail reconstruction is handled by a dedicated decoder.

\paragraph{Speech tokenizer.}
The speech tokenizer converts reference speech waveforms into discrete semantic token sequences.
We adopt the S3Tokenizer V2~\citep{cosyvoice2} as our speech tokenizer. This tokenizer inserts a finite scalar quantization (FSQ) module into a large-scale ASR encoder and is trained end-to-end with an ASR objective, producing a single-codebook token sequence at 25\,Hz that primarily encodes phonemic content rather than acoustic detail. The ASR-supervised training is particularly beneficial for our task, as it makes the token space more discriminative for fine-grained reading differences among kanji.

\paragraph{Backbone LLM.}
We adopt Sarashina2.2-0.5B-Instruct~\citep{sarashina2.2}, a decoder-only LLM pre-trained predominantly on Japanese text, as the backbone LLM and extend its vocabulary by appending 6,561 semantic tokens from S3Tokenizer, together with special tokens ($\texttt{BOS}$, $\texttt{EOS}$, and $\texttt{<|speech\_start|>}$).

The backbone LLM operates in a zero-shot voice cloning setting following the in-context learning paradigm~\citep{valle, tortoisetts}. Given a reference speech prompt $\mathbf{s}^{\mathrm{p}}$ with transcript $\mathbf{x}^{\mathrm{p}}$ and a target text $\mathbf{x}^{\mathrm{t}}$, the input and predicted sequence is formed as:
$$
\underbrace{\texttt{BOS},\;\mathbf{x}^{\mathrm{p}},\;\mathbf{x}^{\mathrm{t}},\;\texttt{<|speech\_start|>},\;\mathbf{s}^{\mathrm{p}}}_{\text{Input}},\;\underbrace{\mathbf{s}^{\mathrm{t}},\;\texttt{EOS}}_{\text{Predicted}}
$$
The model is trained with a teacher-forced causal language modeling objective, computing the cross-entropy loss only over semantic token positions:
$$
\mathcal{L} = -\sum_{t=1}^{T}\log p_{\theta}(\mathbf{s}_{t}\mid\mathbf{x},\mathbf{s}_{<t}),
$$ 
where $\mathbf{x}$ and $\mathbf{s}$ denote a transcript and its corresponding speech sample drawn from the training data, respectively.
Because each training sample consists of naturally contiguous multi-sentence speech, the model learns autoregressive continuation across sentence boundaries, and the inference-time prompt--target arrangement preserves the same sequential structure.

\paragraph{Flow-matching decoder.}
The open-source release of Sarashina2.2-TTS directly adopts the flow-matching decoder from CosyVoice 2~\citep{cosyvoice2}, which uses conditional flow matching (CFM)~\citep{cfm} with a convolutional Transformer UNet architecture. It learns a vector field that transports Gaussian noise to the target mel-spectrogram, conditioned on the semantic token sequence, a reference mel-spectrogram and a speaker embedding. 

\paragraph{Vocoder.}
We adopt HiFi-GAN~\citep{hifigan} as the vocoder. 
It is a generative adversarial network optimized for efficient and high-fidelity speech synthesis, which converts the mel-spectrograms produced by the flow-matching decoder into time-domain waveforms.

Note that these acoustic stage and tokenization components (the speech tokenizer, the flow-matching decoder, and the vocoder) operate independently of the linguistic decisions, meaning they do not affect kanji reading accuracy. Consequently, this work focuses on the training and evaluation of the backbone LLM.

\section{Data Strategy} \label{sec:data}

This section describes the datasets and data preparation strategy used for Sarashina2.2-TTS. We first detail the composition and preprocessing pipeline of our large-scale, multilingual speech–text corpus. We then introduce our targeted synthetic data augmentation pipeline, which complements the large-scale corpus by addressing residual reading errors on infrequent kanji and proper nouns.

\subsection{Training Data} \label{sec:training_data}

\subsubsection{Data Composition}

We train Sarashina2.2-TTS on approximately 361k hours of speech--text paired data: 194k hours of Japanese (53.7\%) across seven domains and 167k hours of English (46.3\%). Note that all audio sources are either licensed or in the public domain. Table~\ref{tab:data_overview} shows the composition by domain.

\begin{table}[h]
\centering
\caption{Training data composition by domain and language.}
\vspace{1mm}
\label{tab:data_overview}
\begin{tabular}{lrrr}
\toprule
Domain & Japanese (h) & English (h) & Total (h) \\
\midrule
Podcast              & 58{,}304  & 106{,}927 & 165{,}231 \\
Audiobook / Narration & 66{,}839  & --        & 66{,}839  \\
Customer Service     & 27{,}864  & --        & 27{,}864  \\
TV / Broadcast       & 21{,}287  & --        & 21{,}287  \\
Public Speech        & 16{,}157  & --        & 16{,}157  \\
Conversation         & 2{,}565   & --        & 2{,}565   \\
Language Learning    & 1{,}035   & --        & 1{,}035   \\
Uncategorized$^\ast$ & --   & 60{,}226  & 60{,}226  \\
\midrule
\textbf{Total}       & \textbf{194{,}051} & \textbf{167{,}153} & \textbf{361{,}204} \\
\bottomrule
\end{tabular}
\vspace{2pt}

{\footnotesize $^\ast$ Sourced from publicly available speech datasets; fine-grained domain labels are not available.}
\end{table}

The Japanese data covers a wide range of speaking styles, from formal narration and broadcast speech to spontaneous conversation and customer-service calls. This domain diversity serves two purposes: (1)~it exposes the model to a wide range of prosodic and stylistic patterns, improving style reproduction in zero-shot voice cloning, and (2)~it broadens the vocabulary and contextual patterns in the transcriptions, helping the model see more diverse kanji usage patterns 
and thereby improve kanji reading accuracy. English data is included for multilingual capability and to handle code-switching inputs where English words appear in Japanese text.

Furthermore, we intentionally balance the ratio of Japanese and English data. This strategy aims not only at robust bilingual speech synthesis but also at enabling cross-lingual prompt conditioning.
By providing a well-balanced mixture of both languages during training, the model acquires the ability to perform cross-lingual voice cloning, successfully transferring a speaker's identity from an English reference prompt to Japanese synthetic speech.

\subsubsection{Preprocessing} \label{sec:preprocessing}

We process all raw audio through a standard preprocessing pipeline that performs  audio standardization, source separation, speaker diarization, voice activity detection (VAD)-based segmentation, speech enhancement for low-quality sources, transcription, language identification, and multi-dimensional quality filtering based on DNSMOS scores and dual-ASR verification. Specifically, we find that Whisper large-v3-turbo~\citep{whisper} can accurately transcribe Japanese speech in most cases but sometimes has repetition errors. To avoid these errors, we use OWSM-CTC-v4~\citep{owsmv4,peng2024owsmctc} as a secondary ASR model for verification, and discard samples with large transcription CER between the two transcriptions. This pipeline produces the 361k hours of clean speech--text pairs used for training.

We note that the semantic tokens used in our system are inherently resilient to minor acoustic artifacts introduced by source separation and speech enhancement~\citep{touchtts}. This robustness allows less stringent filtering criteria and higher data retention rates, which is particularly important for Japanese where large-scale speech data is limited due to strict copyright regulations.

\subsection{Targeted Synthetic Data Augmentation} \label{sec:augmentation}

While large-scale training covers most regular-use kanji-reading mappings, residual errors persist on rare proper nouns, place names, and other infrequent readings. To systematically improve kanji reading accuracy, we design a data augmentation pipeline that generates synthetic training data. This pipeline is built on Pronunciation Steering (PronSteering) model, a text-side pronunciation control mechanism described below.

\subsubsection{Pronunciation Steering (PronSteering)} \label{sec:pronsteering}

For PronSteering model, we introduce two additional special tokens, $\texttt{<|pron\_start|>}$ and $\texttt{<|pron\_end|>}$, into the vocabulary of the backbone LLM described in Section \ref{sec:model}. This mechanism
enables explicit specification of readings by replacing the target kanji in the input text sequence $\mathbf{x}$ with a delimited control fragment $\mathbf{x}_{\mathrm{pron}}$ containing the kana reading and pitch-accent tags:
\begin{quote}
\texttt{<|pron\_start|>} \textit{kana reading + pitch-accent tags} \texttt{<|pron\_end|>}
\end{quote}

\noindent For example, to specify the reading of \ja{``今日''} as \ja{``キョー''} (\textit{kyo}) with a pitch fall after the first mora:
\begin{quote}
Original: \ja{今日はいい天気ですね。} \\
PronSteering: \texttt{<|pron\_start|>}\ja{キョ]ー}\texttt{<|pron\_end|>}\ja{はいい天気ですね。}
\end{quote}

\noindent The pitch-accent tags follow the prosodic symbol method~\citep{kurihara2021prosodic}, using ``\texttt{[}'' for pitch rise and ``\texttt{]}'' for pitch fall. The special token delimiters explicitly isolate the control fragment from the surrounding text, preventing boundary confusion that would arise from naive kana substitution.

\paragraph{Training data construction.}
We apply morphological analysis~\cite{kudo-etal-2004-applying} to extract all noun spans, randomly sample a subset, predict their kana readings and pitch-accent tags using dictionary-based tools, and wrap the predictions in PronSteering format. We construct approximately 4,000 hours of annotated data, primarily from broadcast and audiobook domains where text-speech alignment is most reliable and pitch-accent patterns follow standard norms. We then fine-tuned the Stage~1(Section~\ref{sec:setup}) model with these data to obtained the PronSteering model. 

\paragraph{Applications.}
PronSteering serves two roles in our development pipeline: (1)~as the controlled synthesis mechanism underlying the augmentation pipeline described in Section \ref{sec:augmentation_pipeline}, and (2)~as a user-defined lexicon in internal deployment, where users can pre-register readings for domain-specific terms. The open-source release of Sarashina2.2-TTS does not include PronSteering capability; instead, it benefits from the synthetic data produced by a PronSteering-enabled internal model.

\subsubsection{Augmentation Pipeline} \label{sec:augmentation_pipeline}
The proposed augmentation pipeline is designed to systematically address the model's pronunciation blind spots and rectify reading errors by automatically generating targeted speech–text pairs. It operates through a sequential three-step process as follows:

\begin{enumerate}
    \item \textbf{Generate training sentences:} We use an LLM to generate natural Japanese sentences containing the target kanji in contexts where only the specified reading is valid. The LLM simultaneously produces full-sentence kana annotations with the target kanji's reading explicitly marked. Each candidate undergoes format checking and UniDic-based morphological verification; samples that fail are sent back to the LLM for refinement (up to two iterations).

    \item \textbf{Annotate and synthesize:} For each verified sentence, we locate the target kanji morpheme via dictionary-based tools, extract its reading and pitch-accent pattern, and construct a PronSteering control fragment. A robust matching procedure handles phonological variations common in Japanese such as sequential voicing (\textit{rendaku}) and gemination. Overall, 97.5\% of samples are successfully annotated with prosody tags; the remaining 2.5\% are synthesized with reading control only. Each sentence is then synthesized with diverse speaker prompts to increase data diversity.

    \item \textbf{Quality filtering:} Each synthesized utterance is transcribed by the Kana-ASR model (defined later in Section~\ref{sec:kana_asr}) and checked against the reference pronunciation at both the target kanji segment level and the full-sentence level. Samples exceeding error thresholds at either granularity are discarded.
\end{enumerate}

We apply this pipeline to all 2,136 kanji and their 4,378 readings in the Joyo Kanji List, Japan's officially designated set of regular-use kanji published by the Agency for Cultural Affairs. Because these kanji represent the complete set that a literate Japanese speaker is expected to know, covering all of them ensures a balanced training signal across both common and infrequent readings. In total, we generate approximately 280k synthetic training samples (about 320 hours). After quality filtering with a 95.1\% retention rate, these are mixed into Stage 2 training (Section~\ref{sec:setup}).

\section{Evaluation Framework} \label{sec:eval}

This section describes our evaluation framework, which consists of two key components: the Kana-CER and the Joyo Kanji Yomi Benchmark.

\subsection{Kana-CER} \label{sec:kana_cer}
While standard CER already suffers from orthographic variation in Japanese, it is further complicated by the fact that different ASR models have their own biases toward particular orthographic forms, introducing additional inconsistency. To address this, we propose Kana-CER, which shifts the comparison from the orthographic level to the phonological level. We use an ASR model that outputs kana sequences (Kana-ASR) to transcribe synthesized speech, and compare the result against kana-form reference readings. Because kana is a purely phonological writing system, this eliminates orthographic variation entirely: the kana sequence is identical as long as the pronunciation matches, regardless of which character form the original text uses. 

We apply Kana-CER at two granularities: Kana-CER$_{\text{sent}}$, computed over the full sentence, and Kana-CER$_{\text{kanji}}$, computed only on the kana substring corresponding to a specific target kanji (Section~\ref{sec:benchmark}). We also report standard CER (via Whisper large-v3-turbo) as a reference metric to facilitate comparison with prior work.

\subsubsection{Kana-ASR Model} \label{sec:kana_asr}
We fine-tune Whisper large-v3-turbo~\citep{whisper} to output kana sequences directly from Japanese speech, using the Corpus of Spontaneous Japanese~\cite{maekawa03_sspr}, which is not used in Sarashina2.2-TTS training.
To verify its transcription accuracy, we compute Kana-CER on real recordings from JSUT~\citep{jsut}, obtaining 0.979\%, confirming that the model's own error rate is well below the performance gaps between TTS systems. The Kana-ASR model\footnote{Available at \url{https://huggingface.co/sbintuitions/kana-whisper}} weights is open-sourced together with the Joyo Kanji Yomi Benchmark and the full evaluation scripts to ensure reproducibility.

However, the Kana-ASR model also has its limitations. It focuses purely on phonological information and lacks the language-model-based semantic compensation available in standard ASR systems like Whisper. When synthesized speech exhibits highly colloquial or stylistically extreme pronunciation, Kana-ASR may produce transcription errors even when the pronunciation is intelligible to humans. For example, a slightly reduced pronunciation of \ja{``彼''} in casual speech might be transcribed as \ja{``ハレ''} (\textit{hare}) rather than the correct \ja{``カレ''} (\textit{kare}), while Whisper's language model would recover the correct text. 
To mitigate this, we use a reading-style (i.e., not overly expressive) reference speech and its transcript as the speech prompt $\mathbf{s}^{\mathrm{p}}$ and text prompt $\mathbf{x}^{\mathrm{p}}$ (defined in Section \ref{sec:model}) for all TTS systems during evaluation to encourage more standard pronunciations.

\subsection{Joyo Kanji Yomi Benchmark} \label{sec:benchmark}


Existing TTS evaluation benchmarks compute CER-based metrics at the sentence level. Each test item is a single sentence, and the metric reflects overall pronunciation accuracy of that sentence. This design cannot attribute errors to individual kanji or determine which reading was incorrectly selected. To enable kanji-level error attribution, we construct the Joyo Kanji Yomi Benchmark.

\paragraph{Construction.}
The benchmark covers all 2,136 kanji in the Joyo Kanji List and their 4,378 readings, with 13,095 test samples in total. Some kanji-reading pairs are excluded when the target reading cannot be uniquely disambiguated from the kanji's other readings by sentence context alone. For each kanji-reading pair, we generate multiple natural Japanese sentences using an LLM, requiring that the sentence context uniquely determines the target reading. Each sentence is accompanied by a full-sentence kana annotation in which the reading segment corresponding to the target kanji is marked with \texttt{<>} delimiters for precise localization during evaluation. For example, for the kanji \ja{``審''} with target 
reading \ja{``シン''} in the sentence \ja{``その法案は国会で現在審議中だ。''}, 
the annotation is:
\begin{quote}
\ja{ソノホーアンワコッカイデゲンザイ}\text{<}\ja{シン}\text{>}\ja{ギチューダ。}
\end{quote}
\noindent where the delimiters \texttt{<>} mark the kana substring 
\ja{``シン''} corresponding to the target kanji \ja{``審''}, 
enabling automatic extraction during evaluation.

All sentences and annotations are then reviewed by 35 native Japanese speakers through a three-stage process: (1) correctness check: verifying the target kanji appears with the correct reading and the kana annotation is correctly delimited; (2) disambiguation check: verifying the context uniquely determines the reading; (3) pronunciation correction: reviewing and correcting the full kana annotation. Samples failing either of the first two checks are discarded and regenerated. After verification, three sentences are retained per kanji-reading pair, yielding the final set of 13,095 test samples. 

\paragraph{Evaluation protocol.}
For each test sample, we synthesize the sentence, transcribe it with Kana-ASR into a kana sequence, and use Levenshtein-distance-based alignment to extract the reading substring corresponding to the target kanji. We then compute Kana-CER$_{\text{kanji}}$ between the extracted segment and the reference reading. We use CER rather than exact-match accuracy because Kana-ASR may introduce minor transcription errors (Section~\ref{sec:kana_asr}); CER captures partial correctness and is less sensitive to single-character substitutions than a binary accuracy metric. 
We also report $\text{Kana-CER}_{\mathrm{kanji}}^{\dagger}$, which clips each sample's CER at 1.0 before averaging. Since target kanji segments are short (typically 1–2 characters), minor hallucinations can inflate an individual sample's CER to several hundred percent; clipping prevents these extreme outliers from distorting the overall mean.

\paragraph{Fine-grained error attribution.}
Because each test sample targets a specific kanji-reading pair, the benchmark can tally error rates per reading and pinpoint exactly which readings a system struggles with. This diagnostic capability can directly inform data augmentation strategies by prioritizing readings with the highest error rates for additional training data synthesis (Section~\ref{sec:augmentation}). We demonstrate this analysis in Section~\ref{sec:per_reading}.

\section{Experiments} \label{sec:experiments}

\subsection{Setup} \label{sec:setup}

As described in Section \ref{sec:model}, this work focuses on the backbone LLM. Specifically, following the two-stage training strategy described in Section \ref{sec:data}, we evaluate the performance across three benchmark datasets and compare our system against several baseline LLM-TTS systems.

\paragraph{Training.}

The backbone LLM is a 24-layer Transformer decoder with the extended vocabulary described in Section \ref{sec:model}, initialized from Sarashina2.2-0.5B-Instruct~\citep{sarashina2.2}.
We employ a two-stage training strategy for the backbone LLM as follows:
\begin{itemize}
    \item \textbf{Stage 1 (Pre-training):} 
    Full-parameter training is conducted on the 361k-hour corpus (Section \ref{sec:training_data}) with a constant learning rate of $1 \times 10^{-4}$ to establish the foundational text-to-speech mapping.
    \item \textbf{Stage 2 (Fine-tuning):} 
    Continued training is then performed with a linearly decayed learning rate from $1 \times 10^{-4}$ to $1 \times 10^{-6}$. In this stage, a re-filtered, higher-quality subset of the Stage 1 data is mixed with the targeted synthetic data (Section \ref{sec:augmentation}) to improve the coverage of long-tail readings that are underrepresented in the natural training data.
\end{itemize}
For the open-source release, the decoder and vocoder are directly adopted from CosyVoice 2~\citep{cosyvoice2}. This report focuses on the backbone LLM, which is the only component trained in this work; the acoustic stage components may differ in internal deployments. All experimental results reported in this paper are based on the open-source configuration.

\paragraph{Baselines.}
We compare against four recent LLM-TTS systems that support Japanese: T5Gemma-TTS~\citep{t5gemmatts}, Qwen3-TTS~\citep{qwen3tts}, FishAudio S1-mini~\citep{fishaudio_s1}, and FireRedTTS-2~\citep{fireredtts2}. The latter three are primarily multilingual systems with Japanese as one of the supported languages.

\paragraph{Evaluation datasets.}
We evaluate performance across three distinct benchmarks:
\begin{itemize}
    \item Joyo Kanji Yomi Benchmark: 
    This dataset consists of 13,095 samples covering all 2,136 regular-use kanji. We compute $\text{Kana-CER}_{\mathrm{kanji}}$, $\text{Kana-CER}^{\dagger}_{\mathrm{kanji}}$, $\text{Kana-CER}_{\mathrm{sent}}$, and standard CER as described in Sections 4.1 and 4.2. To encourage standard pronunciations, a fixed reading-style utterance and its transcript are used as the speech prompt $\mathbf{s}^{\mathrm{p}}$ and text prompt $\mathbf{x}^{\mathrm{p}}$ (Section \ref{sec:model}), respectively.
    \item JSUT: 
    We are using the basic5000 subset which includes 5,000 Japanese text–speech pairs with verified kana annotations, evaluated using $\text{Kana-CER}_{\mathrm{sent}}$ and standard CER. To ensure consistent evaluation conditions, we employ the same reading-style configuration for $\mathbf{s}^{\mathrm{p}}$ and $\mathbf{x}^{\mathrm{p}}$ as used in the Joyo Kanji Yomi Benchmark.
    \item CV3-Eval~\citep{cosyvoice3}: 
    This benchmark is adopted for zero-shot speaker similarity evaluation, where we focus specifically on the Japanese subset. The evaluation metric is the speaker similarity (SIM), computed as the cosine similarity between speaker embeddings extracted via CAM++~\citep{campp}. 
\end{itemize}

LLM-TTS systems use sampling-based decoding, so results can vary across random seeds. We run each system with 5 random seeds and report mean $\pm$ standard deviation.

\subsection{Kanji Reading Accuracy} \label{sec:exp_kanji}

We evaluate kanji reading accuracy at two levels: first comparing all systems on aggregate metrics, then analyzing accuracy per reading to identify which readings each system struggles with.

\subsubsection{Overall Results}

Table~\ref{tab:comparative} presents results on the Joyo Kanji Yomi Benchmark and JSUT.

\begin{table}[tb]
\centering
\caption{Results on the Joyo Kanji Yomi Benchmark and JSUT. Best in \textbf{bold}, second-best \underline{underlined}.}
\vspace{1mm}
\label{tab:comparative}
\resizebox{\textwidth}{!}{%
\begin{tabular}{lcccccc}
\toprule
\multirow{2}{*}{System} & \multicolumn{4}{c}{Joyo Kanji Yomi Benchmark} & \multicolumn{2}{c}{JSUT} \\
\cmidrule(lr){2-5} \cmidrule(lr){6-7}
 & Kana-CER$_{\text{kanji}}$ $\downarrow$ & Kana-CER$_{\text{kanji}}^{\dagger}$ $\downarrow$ & Kana-CER$_{\text{sent}}$ $\downarrow$ & CER $\downarrow$ & Kana-CER$_{\text{sent}}$ $\downarrow$ & CER $\downarrow$ \\
\midrule
T5Gemma-TTS              & 13.81 $\pm$ 2.52     & 8.55 $\pm$ 0.45      & \underline{3.69 $\pm$ 1.85}       & 5.68 $\pm$ 2.85      & \textbf{2.80 $\pm$ 0.04} & \textbf{7.63 $\pm$ 0.07} \\
Qwen3-TTS                & 185.89 $\pm$ 105.56  & 21.70 $\pm$ 0.30     & 23.20 $\pm$ 11.53     & 13.26 $\pm$ 3.67     & 15.58 $\pm$ 10.34     & 14.87 $\pm$ 6.47      \\
FishAudio S1-mini        & 33.43 $\pm$ 1.41     & 20.19 $\pm$ 0.20     & 5.46 $\pm$ 0.29       & 6.15 $\pm$ 0.70      & 5.16 $\pm$ 0.05       & 9.03 $\pm$ 0.09       \\
FireRedTTS-2             & 27.82 $\pm$ 0.88     & 16.39 $\pm$ 0.15     & 4.28 $\pm$ 0.06       & \underline{5.32 $\pm$ 0.11}       & 5.26 $\pm$ 0.22       & 9.33 $\pm$ 0.18       \\
\midrule
Sarashina2.2-TTS (Stage 1)   & \underline{11.06 $\pm$ 0.65}     & \underline{6.94 $\pm$ 0.08}       & 4.59 $\pm$ 0.83       & 6.36 $\pm$ 0.63      & 3.04 $\pm$ 0.07       & 8.08 $\pm$ 0.09        \\
Sarashina2.2-TTS (Stage 2)   & \textbf{7.83 $\pm$ 0.70} & \textbf{5.45 $\pm$ 0.10} & \textbf{3.41 $\pm$ 0.77}    & \textbf{5.28 $\pm$ 0.37}       & \underline{2.91 $\pm$ 0.06}       & \underline{8.02 $\pm$ 0.07}        \\
\bottomrule
\end{tabular}%
}
\end{table}

Sarashina2.2-TTS Stage 2 achieves a Kana-CER$_{\text{kanji}}$ of 7.83 and Kana-CER$_{\text{kanji}}^{\dagger}$ of 5.45, substantially outperforming all baselines on the Joyo Kanji Yomi Benchmark. Excluding our own Stage 1 model, the next-best system, T5Gemma-TTS, has a Kana-CER$_{\text{kanji}}^{\dagger}$ of 8.55 which is 57\% higher. Sarashina2.2-TTS also achieves the lowest Kana-CER$_{\text{sent}}$ and standard CER on the Joyo Kanji Yomi Benchmark, indicating the best overall kanji-level and sentence-level accuracy on kanji disambiguation task.

Comparing Stage 1 and Stage 2, the data synthesis pipeline improves all metrics on both benchmarks, \emph{showing the effectiveness of the synthetic data augmentation strategy. }


Across almost all systems and benchmarks, standard CER is consistently higher than Kana-CER$_{\text{sent}}$. For example, Sarashina2.2-TTS Stage 2 shows a gap of 5.11 points on JSUT (2.91 vs.\ 8.02). \emph{This confirms that orthographic variation inflates standard CER and supports the necessity of Kana-CER for Japanese TTS evaluation}. 
Qwen3-TTS stands out as an exception; its frequent hallucinations produce distorted output that disproportionately inflate the Kana-CER. Unlike standard ASR, Kana-ASR lacks the language model compensation necessary to compensate for these errors.

\subsubsection{Per-Reading Analysis} \label{sec:per_reading}

The aggregate metrics in Table~\ref{tab:comparative} summarize overall accuracy, but cannot reveal which readings each system struggles with. Because each test sample in the Joyo Kanji Yomi Benchmark targets a specific kanji-reading pair and marks the corresponding kana segment in the reference annotation (Section~\ref{sec:benchmark}), we can extract the model's actual pronunciation for each target kanji and tally error counts per reading for fine-grained diagnosis. We illustrate this capability using Sarashina2.2-TTS Stage~1 as an example.

Table~\ref{tab:error_examples} shows representative entries from the per-reading error analysis. For each kanji-reading pair, we report the number of mispronounced trials out of 15 (5 seeds $\times$ 3 sentences) together with the actual readings produced by the model and their occurrence counts. The entries are grouped by error severity to illustrate the range of diagnostic information the benchmark provides.

\begin{table}[tb]
\centering
\caption{Examples of per-reading error analysis for Sarashina2.2-TTS Stage~1. Each row shows a kanji-reading pair, the number of mispronounced trials out of 15, and the readings actually produced by the model (with occurrence counts).}
\vspace{1mm}
\label{tab:error_examples}
\begin{tabular}{lllcl}
\toprule
Kanji & Target Reading & Examples & Errors (/15) & Produced Readings \\
\midrule
\ja{六} & \ja{ム} (\textit{mu}) & \ja{六十路}, \ja{六三四} & 15 & \ja{ロクジューロ}(5), \ja{ロクサンシ}(4), ...\\
\ja{坂} & \ja{ハン} (\textit{han}) & \ja{急坂}, \ja{坂路} & 15 & \ja{ザカ}(5), \ja{サカ}(4), ... \\
\ja{従} & \ja{ショウ} (\textit{shou}) & \ja{従容} & 15 & \ja{ジュー}(15) \\

\midrule
\ja{出} & \ja{スイ} (\textit{sui}) & \ja{出納} & 12 & \ja{シツノ}(2), \ja{ズイノ}(2), ... \\
\ja{正} & \ja{マサ} (\textit{masa}) & \ja{正に}, \ja{正夢} & 10 & \ja{セー}(4), \ja{タダ}(4), ... \\
\ja{霜} & \ja{ソウ} (\textit{sou}) & \ja{霜害}, \ja{晩霜} & 10 & \ja{シモ}(5), \ja{ショー}(5) \\
\midrule
\ja{生} & \ja{ショウ} (\textit{shou}) & \ja{生滅}, \ja{一生} & 5 & \ja{セー}(4), \ja{セーミ}(1) \\
\ja{蛇} & \ja{ジャ} (\textit{ja}) & \ja{蛇の目}, \ja{蛇腹} & 4 & \ja{ヘビ}(3), \ja{エビ}(1) \\
\ja{駄} & \ja{ダ} (\textit{da}) & \ja{駄菓子}, \ja{駄作} & 1 & \ja{タ}(1) \\
\bottomrule
\end{tabular}
\end{table}

The error patterns in Table~\ref{tab:error_examples} fall into three interpretable categories. For readings with 15/15 errors, the model consistently falls back to a more frequent reading of the same kanji: \ja{``坂''} is always read as \ja{``サカ''} (\textit{saka}) instead of the rare pronunciation \ja{``ハン''} (\textit{han}), and \ja{``六''} as \ja{``ロク''} (\textit{roku}) instead of \ja{``ム''} (\textit{mu}). These represent readings the model has not learned and are clear targets for data augmentation. For readings with intermediate error counts (e.g., \ja{``正''}-\ja{``マサ''} (\textit{masa}) at 10/15, \ja{``生''}-\ja{``ショウ''} (\textit{shou}) at 5/15), the model succeeds in some sentence contexts but fails in others; examining which contexts trigger correct vs.\ incorrect readings can reveal what contextual cues the model has or has not captured. For readings with very few errors (e.g., \ja{``駄''}-\ja{``ダ''} (\textit{da}) at 1/15), the errors are sometimes attributable to Kana-ASR transcription noise rather than genuine mispronunciation, such as confusing phonetically similar kana pairs (\ja{``ダ''} (\textit{da}) vs.\ \ja{``タ''} (\textit{ta})).

This per-reading analysis can also be applied across systems to compare their reading-level strengths and weaknesses. Table~\ref{tab:per_reading} shows a few representative kanji with both common and rare readings, comparing error counts across all evaluated systems. Common readings are near-universally correct, while rare readings exhibit large cross-system variation, with Sarashina2.2-TTS Stage~2 achieving the lowest error counts in most cases.

\begin{table}[tb]
\centering
\caption{Cross-system per-reading comparison on the Joyo Kanji Yomi Benchmark. Each cell shows the number of mispronounced trials out of 15 (lower is better). Stage 1/2 represents Sarashina2.2-TTS Stage 1/2. Common readings (top row of each group) vs.\ rare readings (bottom rows).}
\vspace{1mm}
\label{tab:per_reading}
\resizebox{\textwidth}{!}{%
\begin{tabular}{llllcccccc}
\toprule
Kanji & Reading & Type & Example & Stage1 & Stage2 & T5Gemma & FireRed2 & S1-mini & Qwen3 \\
\midrule
\ja{事} & \ja{ジ} (\textit{ji}) & common & \ja{事件} & 0 & 0 & 0 & 1 & 0 & 0 \\
\ja{事} & \ja{ズ} (\textit{zu}) & rare & \ja{好事家} & 15 & \textbf{1} & 0 & 15 & 15 & 15 \\
\midrule
\ja{出} & \ja{シュツ} (\textit{shutsu}) & common & \ja{出発} & 0 & 1 & 0 & 0 & 0 & 0 \\
\ja{出} & \ja{スイ} (\textit{sui}) & rare & \ja{出納} & 12 & \textbf{0} & 15 & 14 & 15 & 15 \\
\midrule
\ja{従} & \ja{ジュウ} (\textit{juu}) & common & \ja{従来} & 0 & 0 & 1 & 0 & 0 & 4 \\
\ja{従} & \ja{ショウ} (\textit{shou}) & rare & \ja{従容} & 15 & \textbf{2} & 15 & 15 & 15 & 15 \\
\midrule
\ja{生} & \ja{セイ} (\textit{sei}) & common & \ja{生活} & 0 & 0 & 0 & 0 & 0 & 0 \\
\ja{生} & \ja{ショウ} (\textit{shou}) & common & \ja{一生} & 5 & \textbf{2} & 5 & 5 & 5 & 5 \\
\ja{生} & \ja{オウ} (\textit{ou}) & rare & \ja{生い茂る} & 14 & 14 & 14 & 15 & 14 & 15 \\
\bottomrule
\end{tabular}%
}
\end{table}

\begin{figure}[tb]
\centering
\includegraphics[width=\textwidth]{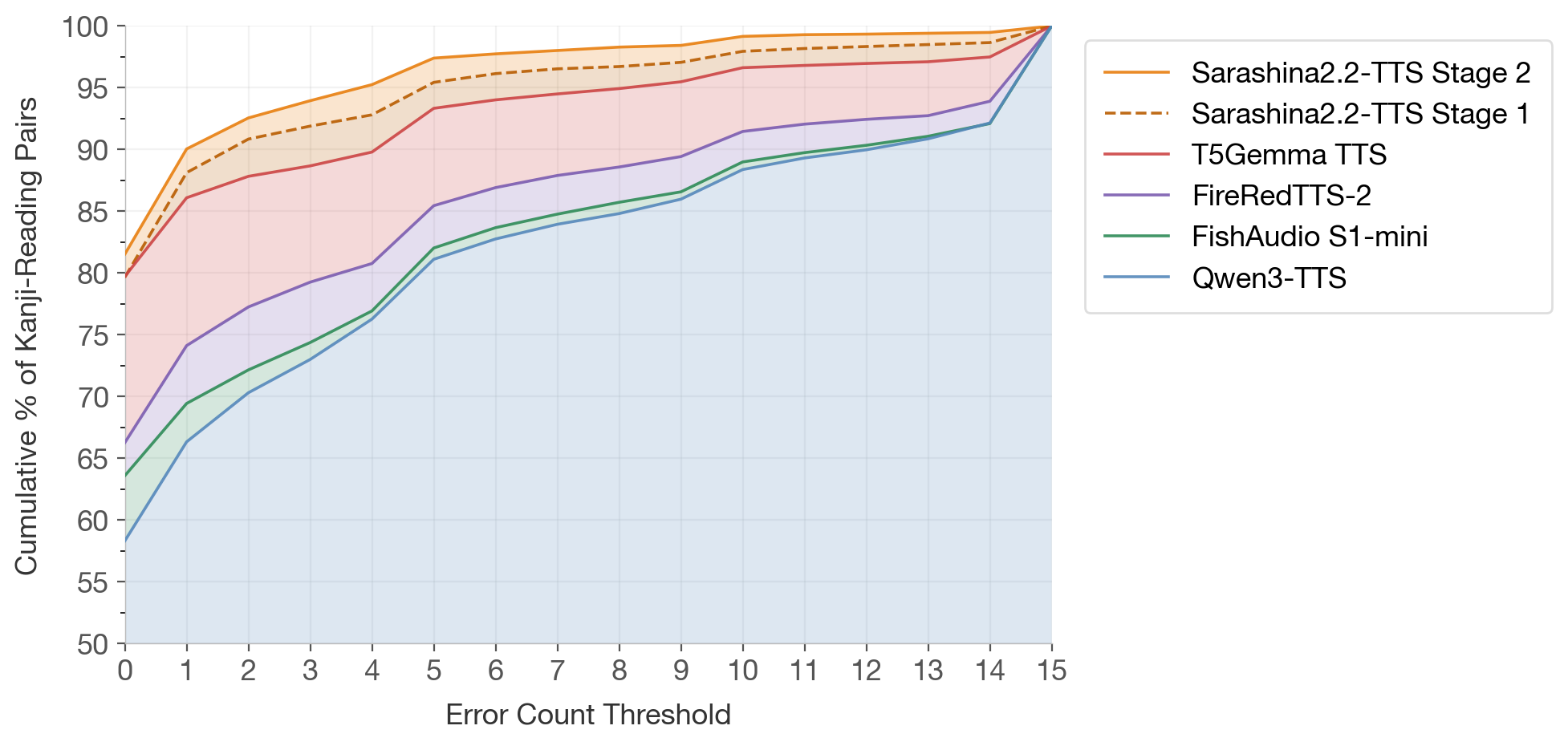}
\caption{Cumulative distribution of per-reading error counts across all 4,378 kanji-reading pairs. Each point shows the percentage of readings with error count $\leq$ the threshold (out of 15 trials). Higher and more left-shifted curves indicate better kanji reading accuracy.}
\label{fig:error_cdf}
\end{figure}

To provide an overall picture of per-reading accuracy across all 4,378 kanji-reading pairs, Figure~\ref{fig:error_cdf} plots the cumulative percentage of readings whose error count falls at or below each threshold. A curve closer to the upper-left corner indicates better performance. Sarashina2.2-TTS Stage~2 leads across the entire range, achieving the highest proportion of fully correct readings (81.5\% at error count = 0), followed by Stage~1 (79.8\%), T5Gemma-TTS (79.7\%), FireRedTTS-2 (66.3\%), FishAudio S1-mini (63.6\%), and Qwen3-TTS (58.3\%). Notably, Stage~1 already ranks second among all systems at every threshold, demonstrating that large-scale multi-domain training alone provides strong kanji disambiguation. Stage~2 further improves upon this through targeted data augmentation, shifting the error distribution toward lower error counts across the board. This confirms that both the large-scale data strategy (Stage~1) and the targeted synthesis pipeline (Stage~2) contribute effectively to kanji reading accuracy on the Joyo Kanji Yomi Benchmark.

In this work, we deliberately synthesize training data for all regular-use kanji readings rather than targeting only error-prone ones, so that the benchmark evaluation remains unbiased (Section~\ref{sec:augmentation}). However, the per-reading error distribution produced by this analysis can directly inform future data strategies by prioritizing the readings with the highest error rates.

\subsection{Cross-Prompt Evaluation} \label{sec:exp_cross_prompt}
In zero-shot TTS, users may provide diverse style or cross-lingual speech prompts as reference. A robust system should maintain consistent pronunciation accuracy regardless of the prompt's language, accent, or speaking style. To evaluate this, we synthesize the JSUT test set with 12 diverse prompts spanning narration, news broadcast, podcast, rakugo, horse-race commentary, and non-Japanese prompts (American, British, and Indian-accented English). We report standard CER (via Whisper large-v3-turbo) rather than Kana-CER, as the stylistic diversity of the prompts can cause Kana-ASR transcription instability (Section~\ref{sec:kana_asr}). We evaluate pronunciation robustness from two perspectives.

\subsubsection{Cross-style Robustness}
Table~\ref{tab:cross_style} reports CER mean and standard deviation across the 9 Japanese prompts, measuring how consistently each system maintains pronunciation accuracy across diverse speaking styles.
Sarashina2.2-TTS Stage 2 achieves superior metrics compared to most baseline models, demonstrating its capacity to handle diverse speaking styles owing to large-scale Japanese pre-training. However, it slightly lags behind FishAudio S1-mini in both mean and standard deviation. 
We hypothesize that this gap partly reflects a limitation of the Whisper large-v3-turbo model used for CER evaluation, rather than a true pronunciation difference: Sarashina2.2-TTS faithfully reproduces highly expressive, acoustically challenging styles such as horse-race commentary, and this high-fidelity prosody cloning may increase transcription difficulty for the ASR model, inflating the apparent CER.

\begin{table}[h]
\centering
\caption{Cross-style evaluation using 9 Japanese prompts. CER is computed via Whisper large-v3-turbo. Each prompt's CER is averaged over 4 seeds; mean and STD are computed across the 9 prompts.}
\label{tab:cross_style}
\begin{tabular}{lcc}
\toprule
System & CER Mean $\downarrow$ & Cross-Style STD $\downarrow$ \\
\midrule
Qwen3-TTS      & 63.11 & 125.00 \\
FireRedTTS-2      & 11.05 & 2.06 \\
FishAudio S1-mini & \textbf{9.44}  & \textbf{0.71} \\
T5Gemma-TTS       & 10.88 & 2.69 \\
\midrule
Sarashina2.2-TTS (Stage 1) & 10.18 & 1.46 \\
Sarashina2.2-TTS (Stage 2)     & \underline{9.55} & \underline{1.32} \\

\bottomrule
\end{tabular}
\end{table}


\subsubsection{Cross-lingual Robustness}
Table~\ref{tab:cross_lingual} compares CER between Japanese and non-Japanese prompts to test whether Japanese pronunciation degrades when the prompt language changes.

\begin{table}[tb]
\centering
    \caption{Cross-lingual evaluation: CER by prompt language group. Degradation = (non-Japanese $-$ Japanese) / Japanese $\times$ 100\%.}
    \vspace{1mm}
    \label{tab:cross_lingual}
    \begin{tabular}{lccc}
    \toprule
    System & Japanese Prompts & Non-Japanese Prompts & Degradation \\
    \midrule
    Qwen3-TTS      & 63.11 & 270.12 & +328.1\% \\
    FireRedTTS-2      & 11.05 & 29.63 & +168.3\% \\
    FishAudio S1-mini & \textbf{9.44}  & 20.14 & +113.4\% \\
    T5Gemma-TTS       & 10.88 & 12.87 & +18.4\% \\
    \midrule
    Sarashina2.2-TTS (Stage 1) & 10.18 & \underline{10.07} & \textbf{$-$1.1\%} \\
    Sarashina2.2-TTS (Stage 2) & \underline{9.55} & \textbf{9.53} & \underline{$-$0.2\%} \\
    \bottomrule
    \end{tabular}
    \end{table}

When prompted with non-Japanese speech, Qwen3-TTS, FireRedTTS-2 and FishAudio S1-mini suffer severe CER degradation (+328\%, +168\% and +113\%), suggesting that their Japanese pronunciation partially relies on acoustic cues from Japanese prompts rather than being fully determined by the input text. T5Gemma-TTS shows moderate degradation (+18\%). Sarashina2.2-TTS is the only system without degradation ($-$0.2\%), indicating that its Japanese pronunciation capability is independent of the prompt language.

We hypothesize that this difference is related to the language balance in training data. In the baseline systems, English is the dominant language: T5Gemma-TTS trains on approximately 170k hours with only about 20k hours of Japanese, and other multilingual baselines similarly allocate the majority of their data to English and Chinese. In contrast, Japanese accounts for 53.7\% of total training hours in Sarashina2.2-TTS. When the training data is dominated by non-Japanese languages, the model may have difficulty separating prompt-side acoustic characteristics from target-side pronunciation decisions, causing the prompt language to bias the synthesized pronunciation.

\subsection{Speaker Similarity}

On the CV3-Ja zero-shot evaluation, Sarashina2.2-TTS achieves the highest SIM scores across both stages (Stage 1: 75.64, Stage 2: 74.75), substantially outperforming all baselines (Table~\ref{tab:sim}). Notably, pronunciation accuracy and speaker similarity do not correlate in a straightforward way among the baselines: Qwen3-TTS ranks second in SIM (69.86) despite poor pronunciation accuracy, while T5Gemma-TTS achieves strong pronunciation but low SIM (50.59).

\begin{table}[h]
\centering
\caption{Zero-shot speaker similarity on CV3-Ja. Best in \textbf{bold}, second-best \underline{underlined}.}
\vspace{1mm}
\label{tab:sim}
\begin{tabular}{lc}
\toprule
System & SIM $\uparrow$ \\
\midrule
T5Gemma-TTS              & 50.59 \\
Qwen3-TTS                & 69.86 \\
FishAudio S1-mini        & 61.38 \\
FireRedTTS-2             & 66.20 \\
\midrule
Sarashina2.2-TTS (Stage 1)  & \textbf{75.64} \\
Sarashina2.2-TTS (Stage 2)  & \underline{74.75} \\
\bottomrule
\end{tabular}
\end{table}

\subsection{Speech Quality}

To verify that the focus on pronunciation accuracy does not come at the cost of speech quality, we evaluate all systems using automatic MOS predictors on the CV3-Ja subset. Table~\ref{tab:mos} reports scores from UTMOS Strong~\citep{utmos}, UTMOS v2~\citep{utmosv2}, DNSMOS~\citep{dnsmos} and DNSMOS P.835~\citep{dnsmos835}.

\begin{table}[h]
\centering
\caption{Automatic MOS evaluation on CV3-Ja. UTMOS represents UTMOS Strong. DNSMOS P.835 is the OVRL score.  Best in \textbf{bold}, second-best \underline{underlined}.}
\label{tab:mos}
\begin{tabular}{lcccc}
\toprule
System & UTMOS $\uparrow$ & UTMOS V2 $\uparrow$ & DNSMOS $\uparrow$ & DNSMOS P.835 $\uparrow$ \\
\midrule
Prompt Speech            & 2.576 & 2.455 & 3.495 & 2.940 \\
\midrule
T5Gemma-TTS          & 2.952 & 2.523 & 3.517 & 3.015 \\
Qwen3-TTS            & \textbf{3.406} & 2.806 & 3.733 & \textbf{3.263} \\
FishAudio S1-mini    & \underline{3.294} & 2.748 & 3.773 & \underline{3.244} \\
FireRedTTS-2         & 2.582 & 2.508 & 3.598 & 3.142 \\
\midrule
Sarashina2.2-TTS (Stage 1)     & 3.184 & \textbf{2.888} & \underline{3.811} & 3.238 \\
Sarashina2.2-TTS (Stage 2)     & 3.174 & \underline{2.877} & \textbf{3.824} & 3.242 \\
\bottomrule
\end{tabular}
\end{table}

Sarashina2.2-TTS achieves the highest scores on UTMOS v2 and DNSMOS, while remaining competitive on the other metrics. Notably, the two stages perform comparably across all quality metrics, confirming that the synthetic data augmentation improves pronunciation accuracy without compromising perceived speech quality. All synthesized systems score above the reference recordings, a known characteristic of automatic MOS predictors that tend to favor the clean, consistent output of TTS over real-world recordings with natural variability.

\section{Conclusion}

We have presented Sarashina2.2-TTS, a Japanese-centric LLM-based TTS system that tackles kanji polyphony---the central challenge of Japanese speech synthesis---through a systematic data strategy and evaluation methodology. On the data side, we train on 361k hours of multi-domain speech and apply targeted data augmentation via PronSteering to cover all regular-use kanji readings. This strategy enables the model to outperform all baselines across all CER-based metrics on the Joyo Kanji Yomi Benchmark while maintaining highly competitive pronunciation accuracy on general sentences. On the evaluation side, we propose Kana-CER to eliminate orthographic variation artifacts in Japanese TTS evaluation and construct the Joyo Kanji Yomi Benchmark for systematic kanji-level error attribution.

Our experiments yield several findings. First, the targeted data augmentation pipeline proves effective: Stage~2's synthetic data covering all regular-use kanji readings improves kanji reading accuracy on both the Joyo Kanji Yomi Benchmark and JSUT over Stage 1, achieving the best kanji-level accuracy among all evaluated systems. Second, the consistent gap between standard CER and Kana-CER across all systems empirically confirms that orthographic variation inflates conventional metrics, supporting the necessity of kana-based evaluation for Japanese TTS. Third, the cross-prompt evaluation reveals that most existing multilingual systems suffer substantial pronunciation degradation under non-Japanese prompts, whereas Sarashina2.2-TTS maintains stable accuracy regardless of prompt language.


Together with this report, we open-source the Sarashina2.2-TTS model weights, the Joyo Kanji Yomi Benchmark, the Kana-ASR model, and the evaluation scripts to facilitate future research in Japanese speech synthesis.

\bibliographystyle{unsrtnat}
\bibliography{references}

\end{document}